# Tidal Limit of Stellar Systems in Two-Power Density Models

M. Alavi*, H. Razmi†

Department of Physics, The University of Qom, 3716146611, Qom, I. R. Iran

*m.alavi@stu.qom.ac.ir

†razmi@qom.ac.ir

**ABSTRACT**

As the generalization of gravitational effects on the point mass systems, we want to study the tidal effect exerted on an extended stellar system using spherical and axisymmetric elliptical models. Considering the Isochrone and Plummer models for a passing extended stellar system, the tidal distance and the equipotential surface are found. The corresponding critical surfaces and maps are plotted for different cases. There are different results some of them may be used in describing stellar systems deformation.

**keywords:** Gravitation -- Galaxies: interactions and evolutions -- Methods: analytical

# 1. INTRODUCTION

As we know, one of the old well-known tidal limit is Roche limit corresponding to the tidal interaction between a planet and its moon (Pugh and Woodworth 2014, Chaisson and Mcmillan 2011). The particles that make up the rings of a planet such as Saturn are often originated from the tidal disruption effect (Zeilik and Gregory 1998). Most of the rings around the planets lie within Roche limits. The discovered comet Shoemaker-Levy9 in 1993, had passed within the Jupiter's Roche limit in 1992 (Kay et al. 2013). Research studies on tidal effects at larger scales than our solar system have been begun in recent years. Recent theoretical and observational advancements reason on this fact that tidal forces can be considered as the main origin of some important phenomena at large galactic and cosmic scales including some inner galactic effects (Weinberg and Blitz 2006), interaction of galaxies (Struck1999), clustering of galaxies (Kravtsov and Borgani 2012), tidal streams (Schneider and Moore 2011) and tidal tails formation (Toomre and Toomre 1972, D' Onghia et al. 2010, Alavi and Razmi 2015). Among these, tail formation is considered as one of the most significant findings from which a number of interesting results about history of galaxies, dark matter distribution and investigation of cosmic halo structure can be found (Renaud et al. 2009). There are also some recent new research results including observational evidences and simulations indicating on direct and indirect tidal effects in capturing of starless planets (Alessandro et al. 2016), chemodynamic evolution of dwarf galaxies in tidal fields (Williamson et al. 2016), the tidal disruption of stars by supermassive black hole binaries (Coughlin et al. 2017) , tidal origin of spiral arms in galaxies (Semczuk et al. 2016), tidal disruption events (Clement Bonnerot et al. 2017, Decker French et al. 2016, Toloba et al. 2016), tidal streams in the Andromeda Galaxy and its Satellite (Choi et al. 2002, Ferguson et al. 2002, Ferguson and Mackey 2016), tidally induced bars of galaxies in clusters (Lokas et al. 2016), the stellar halo and tidal streams of M63 (Staudaher et al. 2015), and tidal debris morphology (Hendel et al. 2015). According to the gravitational property of the tidal forces, at larger than solar system scales, one of the expected effects of the tidal forces is to be related to the mass deformation of the objects. This corresponds to when the tidal forces dominate the self-gravitational forces which can lead to the apparent deformation or even the collapse of the objects. This is why the study and investigation of the limiting distance at which such domination may occur is important. In stellar systems, one usually tries to calculate the so-called tidal radius. Theoretically and ideally, the tidal radius of a globular cluster is equal to the distance at which a star experiences an acceleration towards its host that is equal in magnitude (but opposite in direction) to the acceleration experienced towards the clusters (von Hoerner 1957, King 1962, King 1965). This is the distance up to which the system has a gravitational stability; further from it and due to the tidal effects, the outer parts of the system are separated and its configuration is deformed. There are more rigorous and thorough definitions and derivations of the tidal radius as one in which the restricted three-body problem is used in determining the tidal radius (Binney and Tremaine 2008); of course, both the stellar system and the host galaxy are considered as point masses m and M which are in circular motion around their center of masses. The aim of the restricted three-body problem is to find the path of the test particles in the gravitational field of the two above-mentioned point masses. The approximate solutions of such a problem depends on the motion of the stars in the external parts of the stellar system which is in a circular motion. On the equipotential surfaces, the test particles have no speeds; so, one can find the Hill or Roche radius as

$$r = (\frac{m}{3M})^{1/3} R. \qquad (1)$$

in which R is the distance between the two ideal point masses m and M. This is while in (Innanen et al. 1983), the three-body problem has been studied with an elliptical shape in a power-law mass distribution. The escaped stars from the cluster with interstellar gas can form thin regions named as tidal streams. Some clumps have been observed in the globular cluster Palomar tidal stream (Odenkirchen et al. 2001, Odenkirchen et al. 2003). These are guessed to be due to the tidal epicyclical motion of stars escaping from the star cluster (Kupper et al. 2008). Considering epicyclical motion, the tidal stream structure has been recently analyzed (Just et al. 2009). The tidal radius is also considered in the disruption/deformation of satellites/clusters (Keenan and Innanen 1975, Kravtsov et al. 2004, Baumgardt and Makino 2003), and in the modeling of galaxy formation (Taffoni et al. 2003).

It is useful to have analytical and numerical models for the density and the gravitational potential of galaxies to investigate the dynamical structure of them. The Hubble telescope and some observations show that the luminosity distributions of some galaxies approach a power-law form at small radii (Ferrarese et al. 1994, Crane et al. 1993). These observations excited theorists to construct and develop some new models. Approximating the luminosity of many galaxies show they obey a power law at both the largest and smallest observable radii with a smooth transition at intermediate radii. Numerical simulations of the clustering of dark-matter particles suggest similar structure for the mass density within a dark halo (Navarro J et al. 1995, Navarro J et al. 1996, Navarro J et al. 1997). According to these points, the following density function is of interest in the models under consideration

$$\rho(r) = \frac{\rho_0}{(\frac{r}{a})^\alpha (1+\frac{r}{a})^{\beta-\alpha}}. \qquad (2)$$

Particular cases are known as Dehnen models for $\beta = 4$ (Dehnen 1993) with special ones as Hernquist model for $\alpha = 1$ and $\beta = 4$ (Hernquist 1990), and Jaffe model for $\alpha = 2$ and $\beta = 4$ (Jaffe 1983). The extended versions of Jaffe spherical models to oblate, prolate, and axisymmetric ellipsoidal systems have been introduced as well (Zhenglu 2000, Zhenglu and Moss 2002). In this paper, using Plummer and Isochrone models, we consider a stellar system (using Plummer and Isochrone models) its center moving on a circular orbit with a constant radius of motion. The system is under the influence of the tidal field of a host galaxy its density distribution is assumed to obey some two-power density models. It is clear that we are working with more realized form than the ideal point mass distributions. Our aim is to determine the tidal distances and the equipotential surfaces corresponding to different cases under consideration. The outline of this work is as follows: in section 2, we shortly introduce some selective models for the description of the stellar systems under the influence of the gravitational field of a host system; then, some characteristics of the distributions of the host systems which are considered as some two-power density models (spherical and ellipsoidal ones) are introduced in section 3. Considering the initial conditions as the same as the two preceding sections, the results of the tidal limits corresponding to the stellar system which is in a circular motion and under the influence of the tidal effects of the host system are calculated in section 4. Moreover, the equipotential surfaces passing these limiting distances are found and their corresponding maps are studied. Since the spherical and ellipsoidal distributions under consideration and the overall problem dealing with it are more complicated than the ideal point mass case, the tidal limit deals with a number of new constants and parameters. Section 5 deals with some selective conditions with their corresponding maps. Section 6 consists the results of various scenarios.

## 2. OVERALL DESCRIPTION OF THE PROBLEM

In this section, we introduce the description of the problem including the geometric coordinates of the stellar system and the host galaxy and then a short review of Isochrone and Plummer models governing on the mass distribution of the stellar system.The problem under consideration is about the interaction of a host galaxy and a stellar system. As the stars and all other parts of the stellar system is under gravitational effect of the host galaxy, the main questions correspond to the study of the gravitational potential effects on the stellar system components, investigation of the factors which make extra-gravitational effects dominate self gravitational ones, and finding out the tidal limits due to the second term in the expansion of the gravitational potential function which are of the same values of the self gravitational effects and thus lead to the system deformation.The center of the coordinates system is assumed to be on the center of the orbiting stellar system which is under the influence of the host galaxy in the x-y plane (see Figure1)

In this figure, R is the distance between the centers of the stellar system and the host galaxy; and r is the distance from a typical star in the stellar system to the center of it. As pointed out, all parts of the stellar system are under the gravitational effects of each other and thus it is necessary to introduce the density of the system.

The stellar system density function is assumed to obey the following well-known models:

## 2.1 The Plummer model

Plummer 1911 fitted observations of globular clusters using

$$\rho(r) = \frac{3M}{4\pi b^3}(1+\frac{r^2}{b^2})^{\frac{-5}{2}}, \qquad (3)$$

where M is the total mass of the system and b is the scale length. The density is constant near the center and falls rapidly at large distance; the corresponding gravitational potential is proportional to $r^{-1}$ at large radii and it would be proportional to $r^2$+constant at small radii.A potential with these properties is

$$\Phi = -\frac{Gm}{(r^2+b^2)^{\frac{1}{2}}}. \qquad (4)$$

The potential energy is found as

$$E = \frac{3\pi G m^2}{32b}. \qquad (5)$$

Plummer used these relations to fit observations of some systems.

## 2.2 The Isochrone model

One of the known more realistic models relative to the ideal point mass one is the isochrone model which deals with a potential that is generated by a realistic stellar system for which one can analytically find all orbital equations of motion.

The most general spherical density has been introduced by M. Henon as (Henon 1959a, Henon 1959b, Henon 1960)

$$\rho(r) = M \left[ \frac{3(b^2+r^2)(b+\sqrt{b^2+r^2}) - r^2(b+3\sqrt{b^2+r^2})}{4\pi\sqrt{(b^2+r^2)^3}(b+\sqrt{b^2+r^2})^3} \right]. \quad (6)$$

This is an empirical equation based on observational data for globular stellar systems and, as is seen, it tends to $\frac{bM}{2\pi r^4}$ for large radii.

Henon showed that for a mass distribution (as the same as a point mass), the gravitational potential has this property that all orbits of a given value of energy have the same value of period.

A potential with these properties is

$$\Phi = \frac{Gm}{b + (r^2 + b^2)^{1/2}}. \quad (7)$$

In this model, the most general potential is derived with special role in astronomy because of Henon's potential form with closed form expressions for angle-action coordinates. Binney explain how this property makes the isochrone invaluable for the technique of torus mapping and he believes this model plays a significant role in astrophysics in the coming decade Binney J. (2014).

## 3. SOME TWO-POWER DENSITY MODELS

Before investigating the effective factors on the limiting distance at which the system under consideration is under the influence of, in addition to self gravitational effects, the tidal gravitational forces, it is necessary to introduce the distribution function which describes the gravitational field of the system.

According to observational evidences, luminosity density functions of many galaxies at visible radial distances have power law forms. Simulations corresponding to dark matter candidate particles also predict there is a uniform transition between the smallest and the largest observational distances under the government of some power law models. So, it is important to consider and study such a power law models and their corresponding parameters particularly their power form parameter; this is because, in addition to this consideration that how various values of the parameters affects on the distances at which the system is under the gravitational influence of the host, it is possible to predict their roles on some observations as galactic interactions.

To provide suitable conditions explained about the distribution function of the system, among different types of distributions corresponding to already known models, some certain forms of the distribution (2) known as Jaffe model, Oblate Jaffe model, Prolate Jaffe model, and Hernquist model specified.

In this paper, by studying and comparing the different selective models, one can find the appropriate answers to these questions that, under the same conditions that the system under consideration may have constant distribution (Isochrone or Plummer),Is with changing the parameters, the distances at which the tidal effects are noticeable, have significant differences? Is with changing the spherical distribution corresponding to the Jaffe model for the host galaxy to ellipsoidal one, significant differences are seen for the tidal distances? How do the tidal distances depend on the parameters in the ellipsoidal models? Do these dependences change with changing the distance between the two systems? Does the tidal distance change with changing the distance between the two systems similarly for all the distributions or there are significant changes for some ones and unnoticeable changes for the other ones?

According to these explanations about the distribution are selected:

## 3.1 The Jaffe model

The Jaffe model is introduced by setting $\alpha = 2$ and $\beta = 4$ in (1) (Jaffe 1983):

$$M(r) = 4\pi\rho_0 a^3 \frac{\frac{r}{a}}{1+\frac{r}{a}}. \quad (8)$$

where $\rho_0$ is central density. This is in good agreement with the following luminosity distribution relation which is well-known for many galaxies:

$$\log \frac{I(R)}{I(R_e)} = -3.33 \left[\left(\frac{R}{R_e}\right)^{\frac{1}{4}} - 1\right], \quad (9)$$

where R is the projected radius on the sky, $R_e$ is the effective radius of the isophote enclosing half the light, and I is the brightness (DeVaucouleurs 1948). Application of the Jaffe model to the light distribution of some galaxies yields a good fit with the brightnesses known from other models. The Jaffe distribution function deviates from that of the de Vaucouleurs law at large negative energies (Binney and Tremaine 2008).

### 3.2 The oblate Jaffe model

Now let introduce axisymmetric systems. These models deal with the systems having densities which at large distances decrease as $r^{-4}$ and on the major axis as $r^{-3}$. The exact density function is

$$\rho(r) = \frac{M}{4\pi a} \frac{c^2 d a X^3 + a^2 X^2 (Y^3 + c^2 d) + (3X + 2a) c^2 a Y (Y+d)^2}{X^4 (a+X)^2 Y^3}, \quad (10)$$

where d and c are non-negative and positive constants respectively, and $Y = \sqrt{z^2 + c^2}$, $X = \sqrt{x^2 + y^2 + (Y+d)^2}$ (Zhenglu 2000).

### 3.3 The Prolate Jaffe model

The prolate Jaffe models are extension of the Jaffe's models to axisymmetric systems. These models deal with densities which decrease as $r^{-4}$ at large distances and as $r^{-4}$ on the major axis. The density function corresponding to the prolate Jaffe models are:

$$\rho(r) = \frac{M}{4\pi a} \frac{a^2 Y^2 (X^3 + dX^2 + c^2 d) + daY^3 (X^2 + c^2) + (3Y + 2a) c^2 a X (X+d)^2}{Y^4 (Y+a)^2 X^3}, \quad (11)$$

Where $a$ is length scale, $X = \sqrt{x^2 + y^2 + c^2}$, $Y = \sqrt{(X+d)^2 + z^2}$ and $c$ and $d$ are positive constants. The velocity distributions the prolate models are anisotropic. The projected surface densities for these models cannot be expressed as simply as for the oblate ones Zhenglu and Moss (2002).

### 3.4 The Hernquist model

Consider the density profile

$$\rho(r) = \frac{Ma}{2\pi r (r+a)^3}, \quad (12)$$

where M is the total mass and a is a scale length.

It is similar to Jaffe's model except that $\rho(r) \sim r^{-1}$ as $r \to 0$; so, it has the same behavior as the $r^{1/4}$ law at small radii. There are a number of the properties and distributions of the Hernquist model which are simpler than those of the Jaffe one.

Using (12), the Hernquist model is introduced as:

$$M(r) = 4\pi\rho_0 a^3 \frac{\left(\frac{r}{a}\right)^2}{2\left(1+\left(\frac{r}{a}\right)^2\right)}, \quad (13)$$

which is a volume integral of (12).

## 4. THE METHOD

In this section, we want to determine the distance at which the internal gravitational force of the stellar system becomes equal to the tidal force (the tidal distance) and the equipotential surface corresponding to all the models considered for both the stellar systems and the host galaxies. From physical viewpoints, at larger than these distances, the stability of the system fails and this makes the system deform relative to its former state. To find such a limiting distance, it is enough to equalize the tidal force due to the host system with the gravitational forces keep the stellar system stable.

We are also interested to the equipotential surfaces crossing these limiting distances because by considering such surfaces, we can determine the allowed possible ranges for the motion of the constituent components of the stellar system.

Let consider the system as what is shown in (1). The forces acting on the constituent components of the stellar system are:

$$F = F_{host}(r) - F_{host}(center) + F_{stellar\ system}(r), \quad (14)$$

force due to gravitational potential of the host system.

Considering the nature of the tidal forces as the differential gravitational forces which act on different parts of the stellar system and assuming the distance between the host and the stellar system is greater than the size of the body under the tidal influence, we use following well-known expansion for the gravitational field of the host system ($-\nabla\phi_{host}$):

$$-\frac{\partial\phi_{host}}{\partial x_i}\bigg|_{x} = -\frac{\partial\phi_{host}}{\partial x_i}\bigg|_{x=0} - \sum_j \left(\frac{\partial^2\phi_{host}}{\partial x_i \partial x_j}\right)_{x=0} x_j + \cdots \quad (15)$$

By substituting the equation 15 in 14 and by considering a constant mean value for the angular velocity and by neglecting the Euler acceleration term, the relative acceleration of the constituting parts of the stellar system is found as:

$$a = -\sum \left(\frac{\partial^2\Phi_{host}}{\partial x_i \partial x_j}\right) x_j - \nabla\varphi_{stellar\ system}. \quad (16)$$

According to the assumptions and definitions corresponding to the problem under consideration, to derive the relation for the tidal distance, it is enough to

$$-\sum \left(\frac{\partial^2 \Phi_{host}}{\partial x_i \partial x_j}\right) x_j - \nabla \varphi_{stellar\ system} = 0. \tag{17}$$

As is seen in (Fig 1), the orbital motion of the stellar system has been assumed to be in the x-y plane with an orbital radius of R, and thus the angular velocity is in the z direction.

The expression corresponding to the equipotential surfaces is found by inserting the tidal radius in the potential function.

In what follows, considering the potential functions introduced in sections 3 and 2 for the two systems, the tidal limits corresponding to each model are found one by one.

### 4.1 The Jaffe Model

Using (3), (6), (8), (17), the resulting tidal limits, the tidal distance $x_t$ and the critical surface Crit-Surf, for the Plummer and Isochrone models are found as:

$$x_{t(Plummer)} = [\sqrt[3]{(\frac{M(1+\eta)^2}{4\pi\rho_0(2\eta^4+3\eta^3)})^2} - b^2]^{1/2}, \tag{18}$$

where $\eta = \frac{a}{R}$.

$$Crit - Surf_{(Plummer)}: \ x^2 - \frac{\eta^4+\eta^3}{2\eta^4+3\eta^3} z^2 = \frac{-2(x_t^2+b^2)^{\frac{3}{2}}}{\sqrt{x^2+y^2+z^2+b^2}} + 4b^2 + 2x_t^2. \tag{19}$$

$$x_{t(Isochrone)}: \sqrt{1+\gamma_t^2}[1+\sqrt{1+\gamma_t^2}]^2 = \frac{M(1+\eta)^2}{4\pi\rho_0 b^3(2\eta^4+3\eta^3)}, \tag{20}$$

where $\gamma_t = \frac{x_t}{b}$.

$$Crit - Surf_{(Isochrone)}: \ x^2 - \frac{\eta^4+\eta^3}{2\eta^4+3\eta^3} z^2 = \frac{-2(x_t^2+b^2)^{\frac{1}{2}}\left(b+\sqrt{b^2+x_t^2}\right)^2}{b+\sqrt{x^2+y^2+z^2+b^2}} + 2b^2 + 2b\sqrt{b^2+x_t^2} + x_t^2.$$

$$\tag{21}$$

Already know special point mass system relations are:

$$x_{t(point\ mass)} = \sqrt[3]{\frac{M(1+\eta)^2}{4\pi\rho_0(2\eta^4+3\eta^3)}}, \tag{22}$$

$$Crit - Surf_{(point\ mass)}: x^2 - \frac{\eta^4+\eta^3}{2\eta^4+3\eta^3}z^2 = \frac{-2x_t^3}{\sqrt{x^2+y^2+z^2}} + 3x_t^2. \quad (23)$$

## 4.2 The oblate Jaffe model

Using (3), (6), (10), (17), the resulting tidal limits, the tidal distance $x_t$ and the critical surface Crit-Surf, for the Plummer and Isochrone models are found as:

$$x_{t(Plummer)} = [\sqrt[3]{(\frac{M_{stellar\ system}\ CA^4(A+1)^2}{M_{host}\ (R^2C+3R^2CA+A^2D+DA^3)}a^3)^2 - b^2}]^{1/2}, \quad (24)$$

where $A^2 = \frac{R^2+(d+c)^2}{a^2}, R = \frac{R}{a}, C = \frac{c}{a}$ and $D = \frac{d}{a}$.

$$Crit - Surf_{(Plummer)}: x^2 - \frac{A^{\frac{3}{2}}(D+C)\left(1+\frac{1}{\sqrt{A}}\right)}{2RC+3R^2C\sqrt{A}+DA^{\frac{3}{2}}}z^2 = \frac{-2(x_t^2+b^2)^{\frac{3}{2}}}{\sqrt{x^2+y^2+z^2+b^2}} + 4b^2 + 2x_t^2. \quad (25)$$

$$x_{t(Isochrone)}: \sqrt{1+\gamma_t^2}[1+\sqrt{1+\gamma_t^2}]^2 = \frac{M_{stellar\ system}CA^4(A+1)^2}{M_{host}b^3(R^2C+3R^2CA+DA^2+DA^3)}, \quad (26)$$

$$Crit - Surf_{(Isochrone)}: x^2 - \frac{A^{\frac{3}{2}}(D+C)\left(1+\frac{1}{\sqrt{A}}\right)}{2RC+3R^2C\sqrt{A}+DA^{\frac{3}{2}}}z^2$$

$$= \frac{-2(x_t^2+b^2)^{\frac{1}{2}}\left(b+\sqrt{b^2+x_t^2}\right)^2}{b+\sqrt{x^2+y^2+z^2+b^2}} + 2b^2 + 2b\sqrt{b^2+x_t^2} + x_t^2. \quad (27)$$

$$x_{t(point\ mass)} = \sqrt[3]{\frac{M_{stellar\ system}\ CA^4(A+1)^2}{M_{host}(R^2C+3R^2CA+DA^3+A^2D)}a^3}, \quad (28)$$

$$Crit - Surf_{(point\ mass)}: x^2 - \frac{A^{\frac{3}{2}}(D+C)\left(1+\frac{1}{\sqrt{A}}\right)}{2RC+3R^2C\sqrt{A}+DA^{\frac{3}{2}}}z^2 = \frac{-2x_t^3}{\sqrt{x^2+y^2+z^2}} + 3x_t^2. \quad (29)$$

## 4.3 The Prolate Jaffe Model

Using (3), (6), (11), (17), the resulting tidal limits, the tidal distance $x_t$ and the critical surface Crit-Surf, for the Plummer and Isochrone models are found as:

$$x_{t(Plummer)} = [\sqrt[3]{(\frac{ME^{\frac{3}{2}}F^2(1+F)^2}{4\pi\rho_0(R^2\sqrt{E}+2R^2\sqrt{E}F-F(1+F)(E-R^2))})^2 - b^2}]^{1/2}, \quad (30)$$

where $E = \frac{c^2+R^2}{a^2}$ and $F = \frac{d}{a} + \sqrt{E}$.

$$Crit-Surf_{(Plummer)}: \quad x^2 + \frac{E^{\frac{3}{2}}(1+F)}{R^2\sqrt{E}-(1+F)(2R^2\sqrt{E}+R^2F-EF)}z^2 = \frac{-2(x_t^2+b^2)^{\frac{3}{2}}}{\sqrt{x^2+y^2+z^2+b^2}} + 2x_t^2 + 4b^2. \quad (31)$$

$$x_{t(Isochrone)}: \sqrt{1+\gamma_t^2}[1+\sqrt{1+\gamma_t^2}]^2 = \frac{M E^{\frac{3}{2}}F^2(1+F)^2}{4\pi\rho_0 b^3(R^2\sqrt{E}+2R^2\sqrt{E}F-F(1+F)(E-R^2))}, \quad (32)$$

$$Crit-Surf_{(Isochrone)}: x^2 + \frac{E^{\frac{3}{2}}(1+F)}{R^2\sqrt{E}-(1+F)(2R^2\sqrt{E}+R^2F-EF)} = \frac{-2(x_t^2+b^2)^{\frac{1}{2}}\left(b+\sqrt{b^2+x_t^2}\right)^2}{b+\sqrt{x^2+y^2+z^2+b^2}} + 2b^2 +$$

$$2b\sqrt{b^2+x_t^2} + x_t^2. \quad (33)$$

$$x_{t(point\ mass)} = \sqrt[3]{\frac{M E^{\frac{3}{2}}F^2(1+F)^2}{4\pi\rho_0(R^2\sqrt{E}+2R^2\sqrt{E}F-F(1+F)(E-R^2))}}, \quad (34)$$

$$Crit-Surf_{(point\ mass)}: \quad x^2 + \frac{E^{\frac{3}{2}}(1+F)}{R^2\sqrt{E}-(1+F)(2R^2\sqrt{E}+R^2F-EF)}z^2 = \frac{-2x_t^3}{\sqrt{x^2+y^2+z^2}} + 3x_t^2, \quad (35)$$

### 4.4 The Hernquist model

Using (3), (6), (13), (15), the resulting tidal limits, the tidal distance $x_t$ and the critical surface Crit-Surf, for the Plummer and Isochrone models are found as:

$$x_{t(Plummer)} = \sqrt{(\frac{M\ \alpha(1+\alpha)^3}{2\pi\rho_0(1+3\alpha)})^{2/3} - b^2}, \quad (36)$$

where $\alpha = \frac{R}{a}$.

$$Crit-Surf_{(Plummer)}: \quad x^2 - \frac{2(1+\alpha)}{1+3\alpha}z^2 = \frac{-2(x_t^2+b^2)^{\frac{3}{2}}}{\sqrt{x^2+y^2+z^2+b^2}} + 2x_t^2 + 4b^2. \quad (37)$$

$$x_{t(Isochrone)}: \sqrt{1+\gamma_t^2}[1+\sqrt{1+\gamma_t^2}]^2 = \frac{M\ \alpha(1+\alpha)^3}{2\pi\rho_0 b^3(1+3\alpha)}, \quad (38)$$

$$Crit-Surf_{(Isochrone)}: \quad x^2 - \frac{2(1+\alpha)}{1+3\alpha}z^2 = \frac{-2(x_t^2+b^2)^{\frac{1}{2}}\left(b+\sqrt{b^2+x_t^2}\right)^2}{b+\sqrt{x^2+y^2+z^2+b^2}} + 2b^2 + 2b\sqrt{b^2+x_t^2} + x_t^2. \quad (39)$$

$$x_{t_{(point\ mass)}} = \sqrt[3]{\frac{M\,\alpha(1+\alpha)^3}{2\pi\rho_0(1+3\alpha)}}, \quad (40)$$

$$Crit-Surf_{(point\ mass)}: \quad x^2 - \frac{2(1+\alpha)}{1+3\alpha}z^2 = \frac{-2x_t^3}{\sqrt{x^2+y^2+z^2}} + 3x_t^2. \quad (41)$$

As is clear from the definition of the tidal forces and as what we know about these forces at small scales (The Earth-Moon system), the effects of such forces, irrespective of the initial form of the system, are seen in changing the shape and the symmetry of the system. The different tidal limits derived for the different models of the host system reason on that the limits depend on how one chooses the parameters and constants in the model; thus, to estimate better results it is necessary to have more careful choices. As an example, assuming the stellar system model is fixed, for the Plummer model, the relations (18), (24), (30) and (36) shows that we cannot disregard the role of choosing an appropriate distribution with suitable constants.

Comparison of the obtained relations for the different stellar system distributions shows that choosing an appropriate model for the distribution function of the system under the influence of the tidal effects is also important; this is due to the modifications we have considered than what is considered for the simple ideal point mass model.

As an example, in the relations (18), (20), (24), (26), (30), (32), (36), (38) and (38) corresponding to the models Plummer and Isochrone respectively, it is seen that the tidal distances have new dependences on scale lengths of the models under consideration while there is no such a dependence for the case of point mass system with no any scale length (22), (28), (34) and (40). Moreover, comparing the two relations (18) and (20), it is found that there are two different functional forms of dependences in Isochrone and Plummer models. There is a question that if these differences are seen in the equipotential surfaces too. The comparison of the equations corresponding to the equipotential surfaces (e.g. the relations (19), (21) and (23) ) shows that the answer to the question is positive. In what follows, for better understanding of these changes and dependences, let consider the corresponding results graphically. This is because, by graphical analysis and finding out the necessary information about the tidal limits behaviors at specified values of the constants and parameters, one can find more accurate information about the apparent configuration, and the deformation of a system and the origins of its shape evolution and also its environmental structures.

## 5. THE RESULTING MAPS

Now, let see the resulting maps corresponding to the analytically calculated tidal limits in the previous section. The constants are considered as G=1 and $M_{stellar\ system} = 1$ The three - dimensional equipotential surfaces are plotted for different values of the scale length parameter and $x_t = 1$. The projections of these surfaces in the x - y plane and x - z plane are plotted too. Under the condition that the tidal force exerted on the system to be greater than its self gravitating force, for suitable scale parameters, the stars are possible to escape from the equipotential surfaces which aren't completely closed, this may result in the formation of structures as tidal streams (Figures 2 and 3 for scale parameter of the value of 0.8). The tidal streams, as some important structures under consideration of researches in recent years (Fellhauer M et al. 2007; Montouri M et al. 2007; Besla G et al. 2010), are longitudinal structures already observed around some stellar systems as Palomar Assuming these are originated from external tidal effects and considering the definition of the equipotential surfaces passing through tidal limits, it is enough an equipotential surface has an outer aperture (as two exit tracks seen in the Fig 2 and Fig 3 for the scale parameter value of 0.8) through which some parts can exit from the gravitational bound of the system to its external area and then such a linear structure is formed.

Clearly, the equipotential surfaces and thus the possible places for the motion of stars vary for different values of the scale parameter. ed for two different values of and have been plott ᵗFigures .zAs is seen from the corresponding equations, to consider the effect of different models for the host potential, it is enough to check different values of the z component; the resulting maps under consideration have been plotted based on this point. Choosing appropriate values of the parameters appeared in any model based on real observational data may lead to a variety of results .For the axisymmetric models, the cases similar to elliptical and lenticular galaxies have considered based on the Table 1.

These values, depending on different possible choices of the parameters appeared in the models under consideration, corresponds to the various selective models of the host system. Comparing the relations 19, 21, and 23, as is expected, the left hand sides of the equations which correspond to the distribution of the same host system are equal to each other and only the right hand sides of the equations change by selecting different stellar system models. Also, the left hand sides of the relations 19, 25, 31 and 37 show that the role of choosing different models of the host system appears in the coefficient z; so, the influence of changing the host system distribution can be determined by plotting the corresponding figures for various values of the coefficient z. Of course, it is necessary to choose suitable values of this coefficient for different models of under consideration. For example, in the relation 19 which corresponds to the equipotential surface of the Jaffe model, this coefficient has been appeared as $-\frac{\eta^4+\eta^3}{2\eta^4+3\eta^3}$ which is a function of $\eta$ and should be chosen based on the appropriate value of $\eta$ in accordance of the conditions of the observed system. For ellipsoidal models, these coefficients depend on more variables and for finding out more exact results it is necessary to fixing these parameters. For example, for the oblate Jaffe model, all two parameters C and D should be chosen and substituted suitable to the system under study. As is seen, all the parameters considered in our calculation have their own roles in achieving better results for the tidal limits of a system.

In order to have an outward overview of the ellipsoidal systems which are under consideration in this paper and also to see how the scale parameters and other variables affect on the host system distribution relative to the spherical systems, the density contours corresponding to the oblate Jaffe and the prolate Jaffe models have been shown in Fig 6 and Fig 9. As is seen from the figures, by increasing the value of c, while the value of d is constant, the system distribution becomes more symmetric and closer to the spherical case. The tidal limits variations relative to different values of the distances have been plotted in Fig 7 and Fig 10. Fig 7 corresponds to when the host system distribution is considered as oblate Jaffe model. In these figures, for any constant values of c and d, the various behavior of the tidal limit of the system relative to different values of the distances have been presented graphically for the models under study in this paper (point mass and Plummer). As is seen from the figures, for the case the scale parameter b is smaller than the other ones, as is expected, the resulting map is closer to the point mass case. Similar cases of studies have been considered for a system with the distribution prolate Jaffe in Fig 10; in both figures, it is seen that the tidal limit is increased by increasing the distance.

Variations of the tidal limit relative to the parameters c and d for stellar systems with larger sizes than the star clusters have been represented in Fig 8 and Fig 11.

For the distributions considered in Table 1, Figures 7 and 10 show how the tidal limit depends on the two systems distance when the stellar system is as large as the star clusters.

According to the distributions in figure 6 and by comparing the plotted maps in the Fig 7, it is found that the values of the tidal limits are greater for when the host ellipsoidal system is more symmetric. The other important unavoidable point which can be notified through considering the resulting maps is the role of the distance between the two systems in determining the tidal limits. The resulting maps in Fig 7 and Fig 10 are not completely equivalent because although both distributions correspond to a non-spherical host system, as is seen in the density contours, the ellipsoidal cases have been introduced in two oblate and prolate models.

Based on the relations 24 and 30 and this fact that the tidal limits values aren't independent of the details of the host galaxy distribution, it is found that these limits, in addition to their dependence on R which was considered in Fig 7 and Fig 10, depend on how one chooses the values of the constants c and d.

Fig 8 and Fig 11 have been represented to investigate how do various values of these two constants affect on the tidal limits. Fig 8 shows this limit changes for a stellar system with Plummer distribution model (with scale lengths 0.6 and 0.06) and under the influence of the distribution oblate model due to changing different values of these two constants. This figure consists of three maps where the information corresponding to the chosen values of each one have been represented at the above of it. With increasing the two system distance in terms of smaller values of the constants, the greater values of the tidal limits are obtained. The overall result from different cases checked in this figure is that the tidal limits are under the influence of working with the Plummer stellar system distribution model and choosing its scale length value.

Fig 11 shows this limit for a stellar system in Plummer model (with scale lengths 0.6 and 0.06) and under the influence of the non-spherical prolate model as a function of these two constants. The two system distance influence on the results (for two values of R) has been shown in Fig 11 too.

For constant values of the distances, changing the scale length of the stellar system distribution has direct effect on the tidal limit with the same constants. Comparison of Fig 8 and Fig 11, by considering the distributions in two non-spherical models and the tidal limits dependence in them, shows that the resulting behaviors and values of the tidal limits are different.

Fig 12 shows the density contours corresponding to the two spherical distributions of the Jaffe and Hernquist models in terms of the arbitrary scale parameter equals to one. As is seen, in terms of the same scale parameter (equals to one), the Jaffe model describes a more uniform system than the Hernquist model and thus it is a suitable model for such a distribution. It should be mentioned that this fixed value of the scale parameter (equals to one) and also the distribution under consideration (in Fig 12) have been chosen only as an example for better intuitive understanding of the system, if not, it is necessary to have exact information about the host system. The maps in Fig 13 show how the tidal limits change with changing the distances in spherical Jaffe and Hernquist models. In these maps, as the same as those ones in Fig 7, the distribution of the system under the influence of the tidal effects has been considered for Plummer model with the scale lengths 0.06, 0.6 and 0.3.

Recently, NGC 2974 has been studied (Canizares et al. 1987, de Vauouleurs et al. 1991, DemoulinUlrich et al. 1984, Forman et al. 1985). The total mass of NGC2974 is estimated as $M \sim 1.2 * 10^{11} M_\odot$ (Cinzano and van der Marel 1994). Zhenglu Jiang 2000 have obtained a good fit of an oblate Jaffe model by mimicking the image and ellipticity of this galaxy, and comparing the Gauss-Hermite fit parameters of the model with the corresponding observations of the galaxy. The derived total mass of the model is close to the total mass of this galaxy. The model of Aaronson et al. 1982 gives a= 0.06 kpc. Then the total mass M of the model, given by c = 8.43a and d = 15a, can be calculated as $M = 1.2 * 10^{11} M_\odot$. This mass is very close to (Cinzano and van der Marel 1994).

Using (24) the tidal distance for the stellar system as the Plummer model, given by $M = 3 * 10^4 M_\odot$, b=2pc, in NGC2974 is found 5.58pc and using (28), the resulting distance for the point mass system will be 7pc.

Supposing this system is as an ideal point mass, and calculating the radius based on classical (1), the amount equals 3.3pc.

## 6. CONCLUSION

we have generalized the tidal effect of a host galaxy on stars of a stellar system from already studied point mass systems to some well-known extended distribution models. The tidal limits, including the tidal distances and the equipotential surfaces for a number of models have been determined both analytically and graphically According to this fact that the tidal forces originated from differential gravitational forces, they can deform the stellar systems and thus they may be considered as possible candidates for secondary structures formed around the stellar systems. In more details, these types of tidal structure formations depend on a number of factors among them are the distribution function of the stellar system some ones we have considered in this paper, the distribution function of the system to be tidally formed around the stellar system, some parameters and constants corresponding to the host system some ones studied in this paper and how one may choose suitable values of such parameters and constants, and the tidal distances corresponding to the systems under study. Based on these considerations, the different results found here can be used in studying stellar systems deformation and the formation of some structures as tidal streams. Although there are a number of idealizations, for those observational data which are relatively in agreement with the assumption of the circular motion of the system, by suitable choice of the density distribution, it is possible to get more exact and real results.

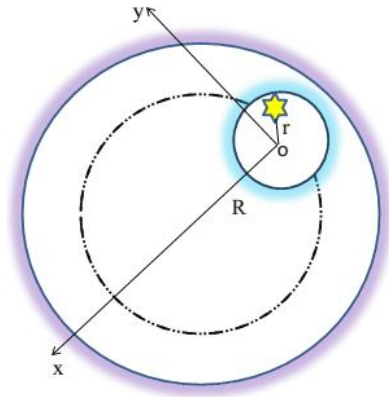

**Figure 1**: An illustration of the problem

| sequence | a | c | D |
|---|---|---|---|
| Between E6 and E7 | 1 | 0.5 | 1 |
| Between E5 and E6 | 1 | 1 | 1 |
| Between E1 and E2 | 1 | 5 | 1 |

Table 1

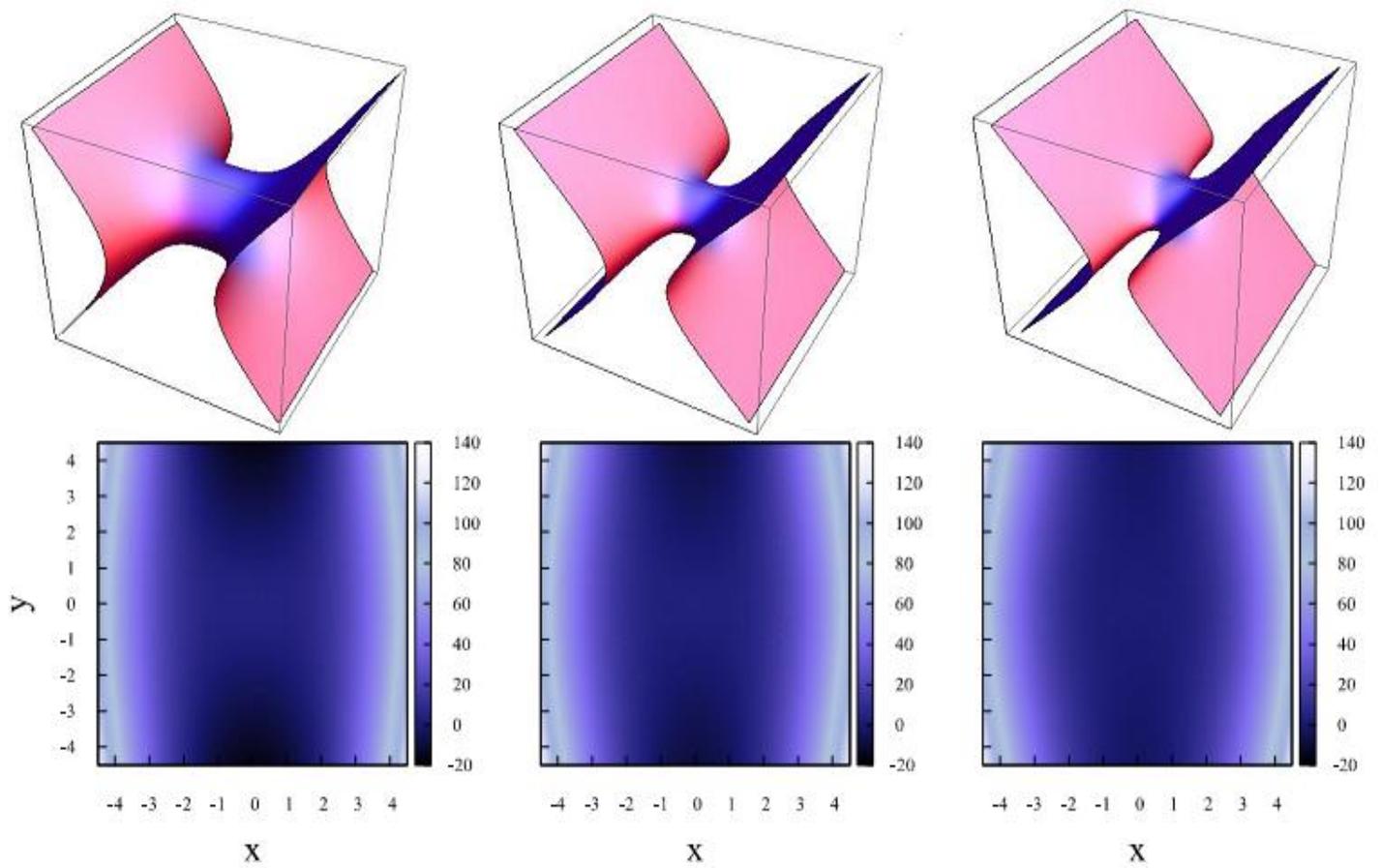

**Figure 2**: The first line 3 maps are the equipotential surfaces corresponding to Isochrone model. The second line 3 maps are the two-dimensional surfaces in *x-y* plane. The relative value of scale parameter to the tidal limit from right to left is 0.06, 0.6, and 0.8 respectively.

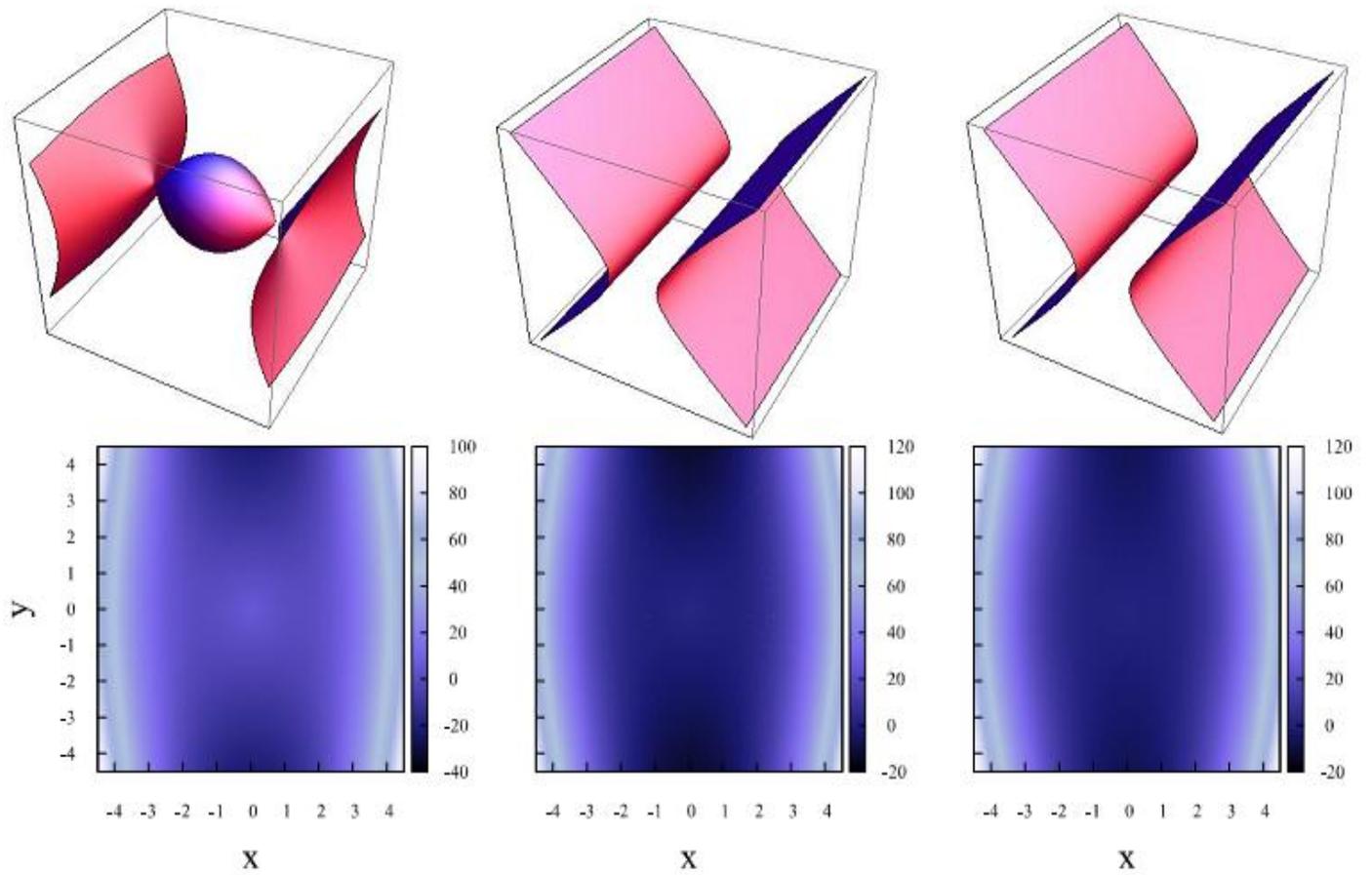

**Figure 3**: The first line 3 maps are the equipotential surfaces corresponding to Plummer model. The second line 3 maps are the two-dimensional surfaces in *x-y* plane. The relative value of scale parameter to the tidal limit from right to left is 0.06, 0.6, and 0.8 respectively.

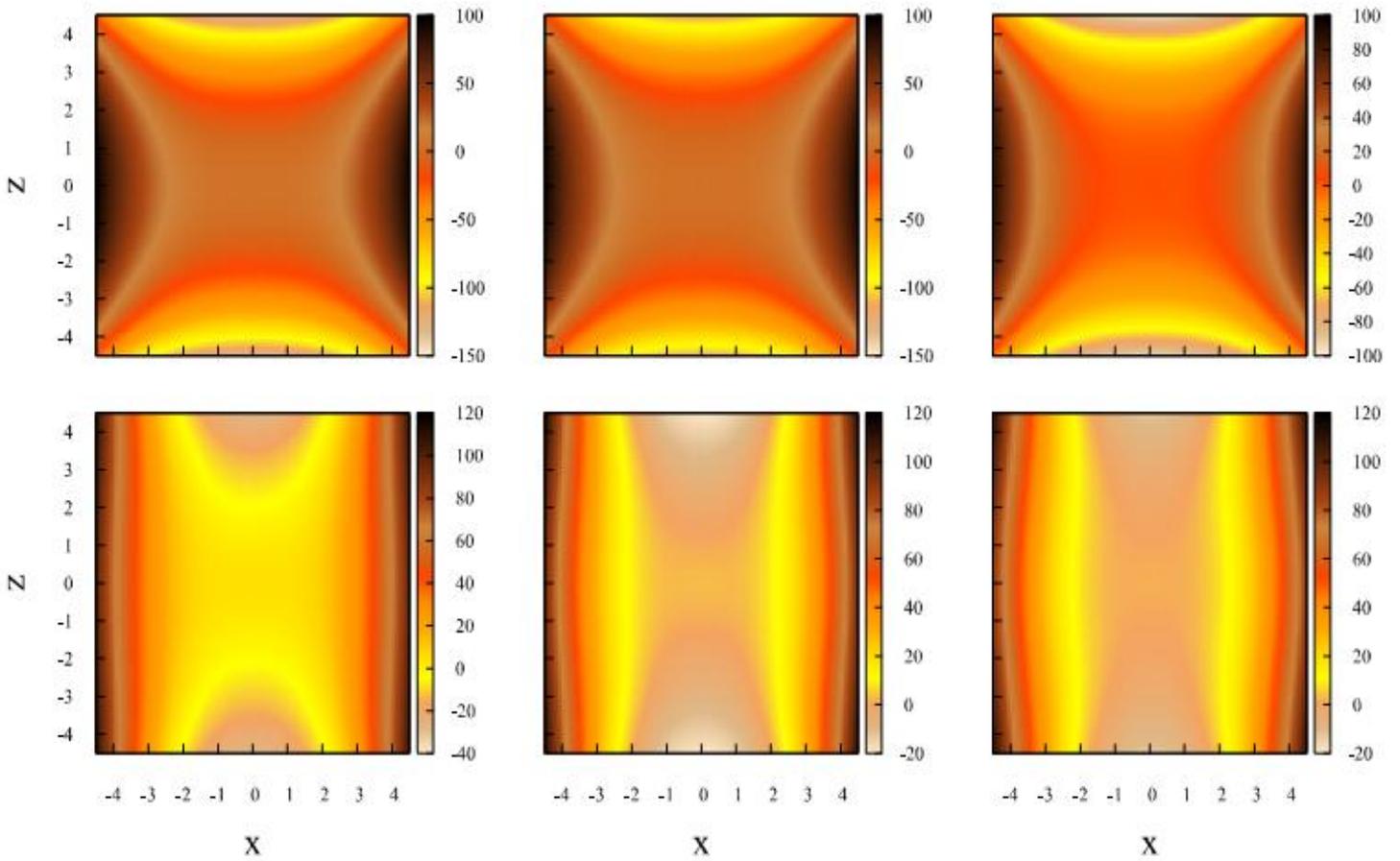

**Figure 4**: The first line 3 maps are the equipotential surfaces corresponding to Isochrone model in *x-z* plane. The second line 3 maps are the equipotential surfaces considering the effect of the host galaxy model. The relative value of scale parameter to the tidal limit from right to left is 0.06, 0.6, and 0.8 respectively.

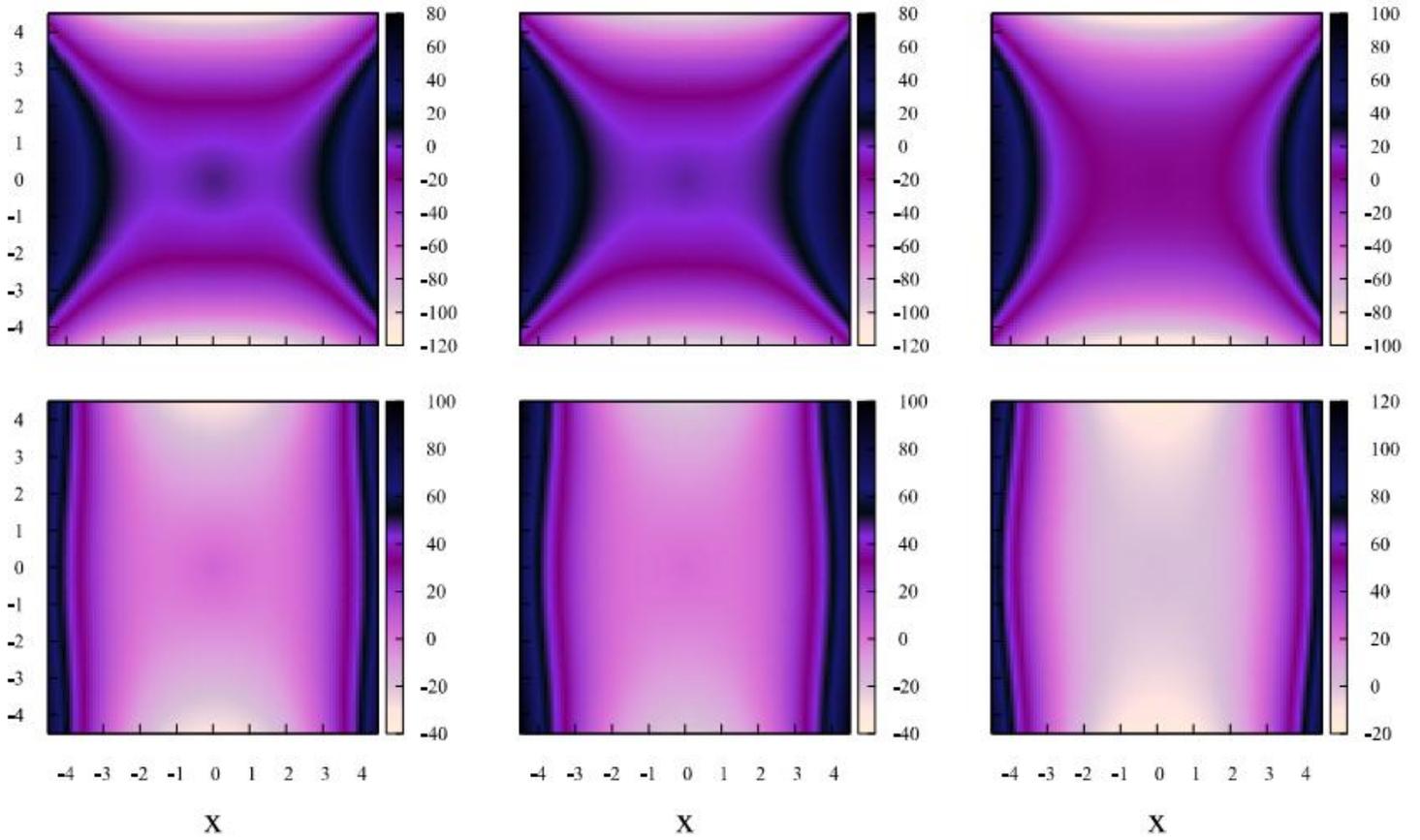

**Figure 5**: The first line 3 maps are the equipotential surfaces corresponding to Plummer model in *x-z* plane. The second line 3 maps are the equipotential surfaces considering the effect of the host galaxy model. The relative value of scale parameter to the tidal limit from right to left is 0.06, 0.6, and 0.8 respectively.

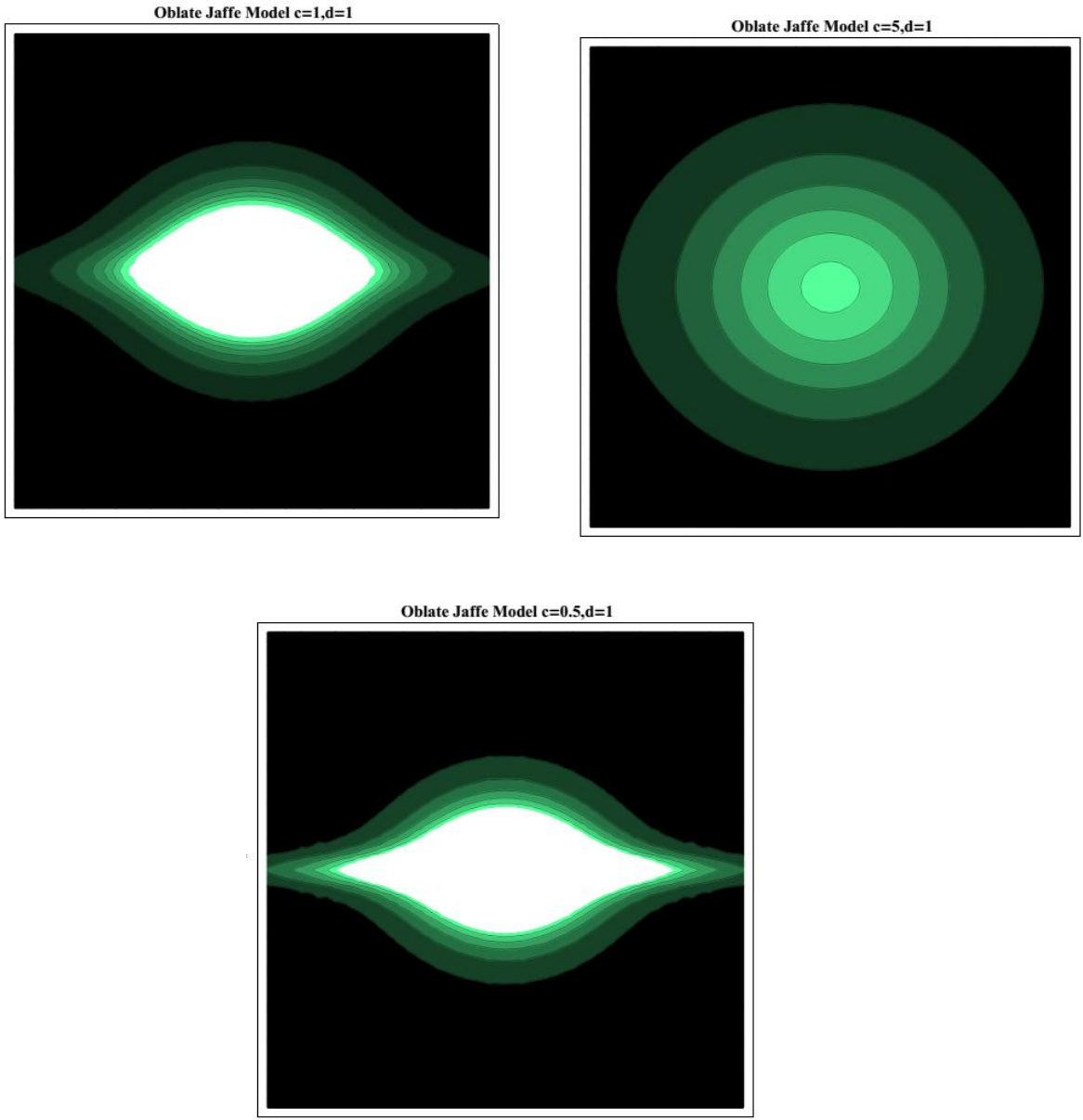

**Figure 6**: The density contours corresponding to Oblate Jaffe model in terms of different constants when the scale parameter is equal to 1.

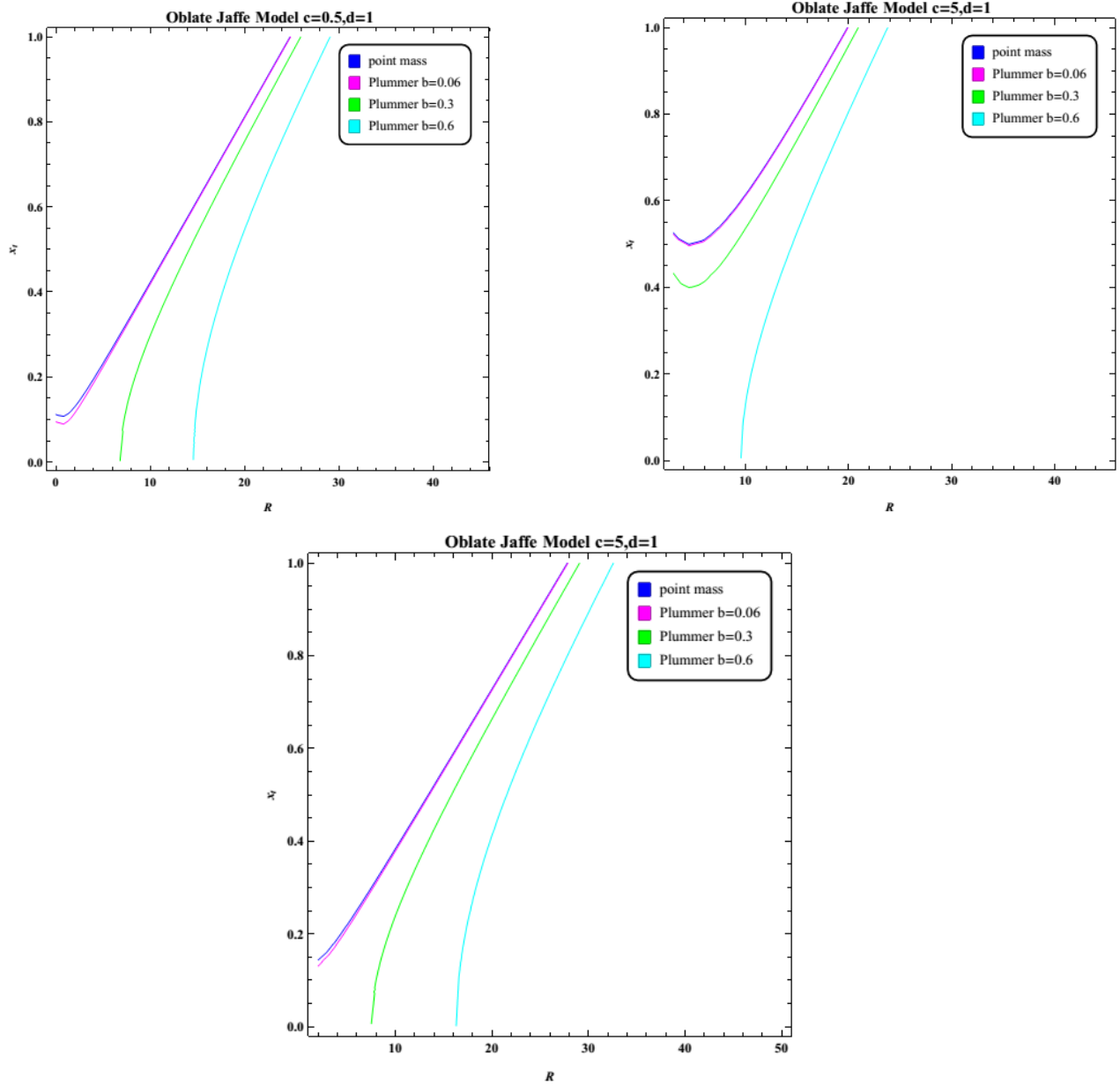

**Figure 7**: These curves show variation of tidal radius relative to the two systems distance. The galaxy model is based on the distributions shown in figure 6; the stellar system density is as the point mass and the Plummer models for scale lengths 0.06, 0.3, and 0.6.

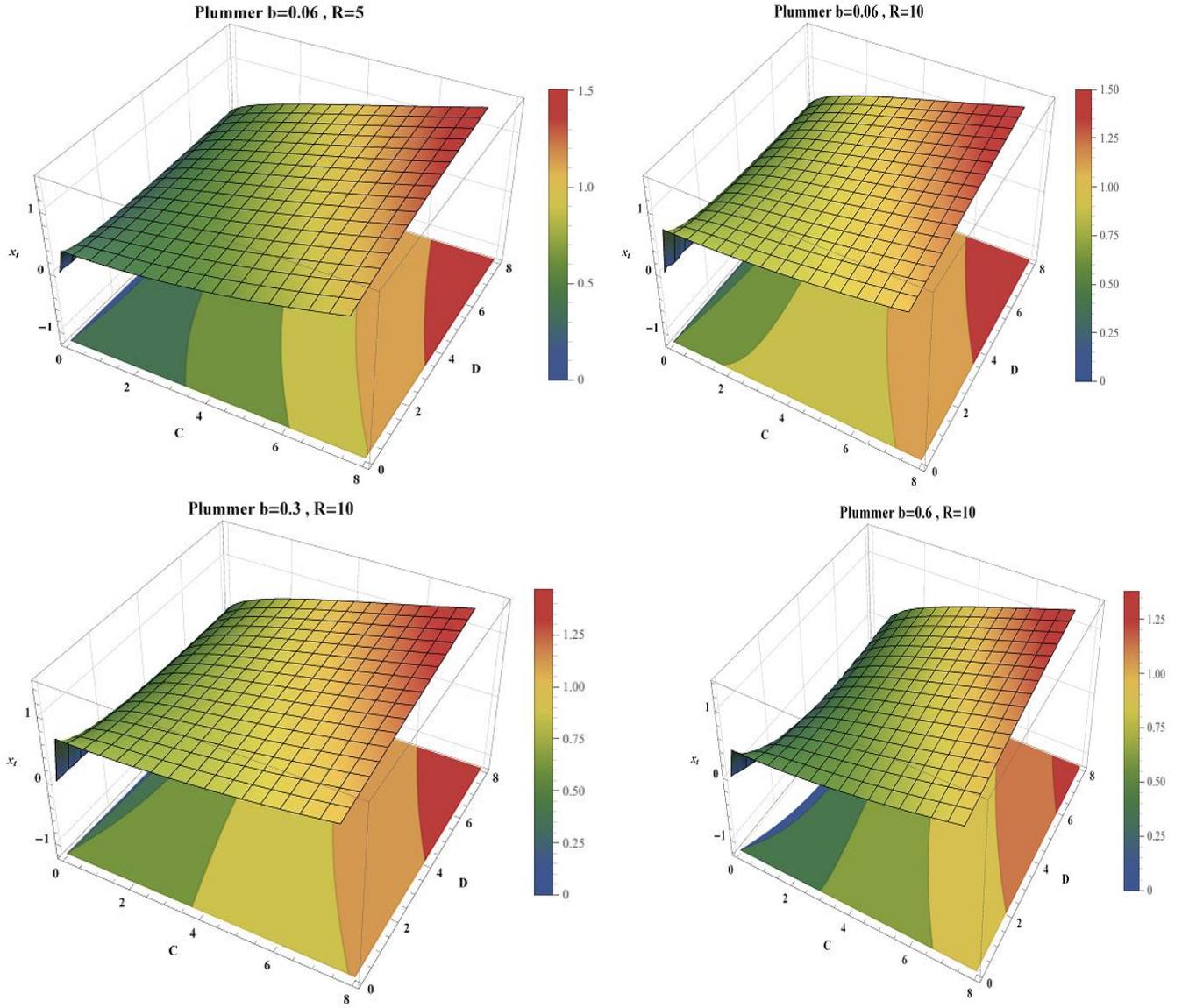

**Figure 8**: These curves show variation of tidal radius in terms of the two constants *C* and *D* (equation 19) for dimensionless distances 10 and 5. The galaxy model is an oblate model and the stellar system density model is the Plummer one for scale lengths 0.06, 0.3, and 0.6.

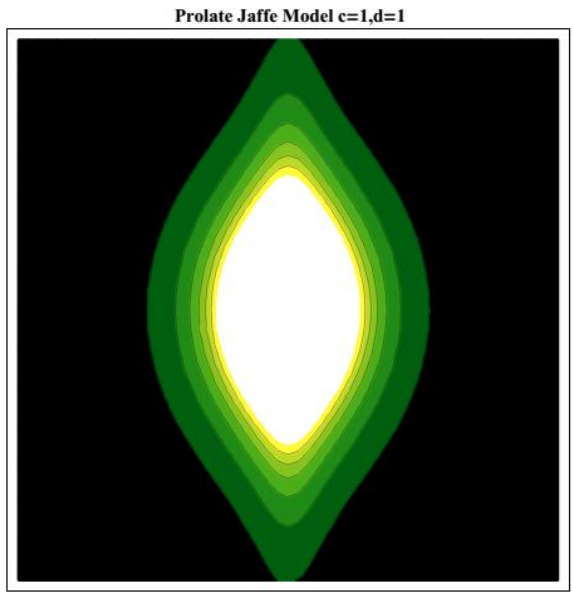
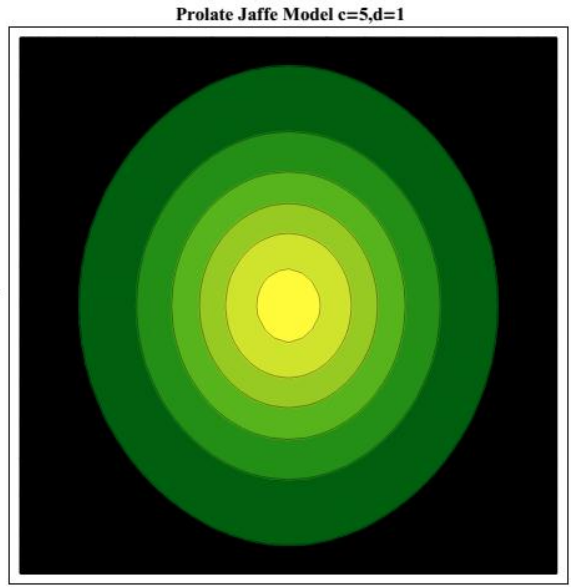
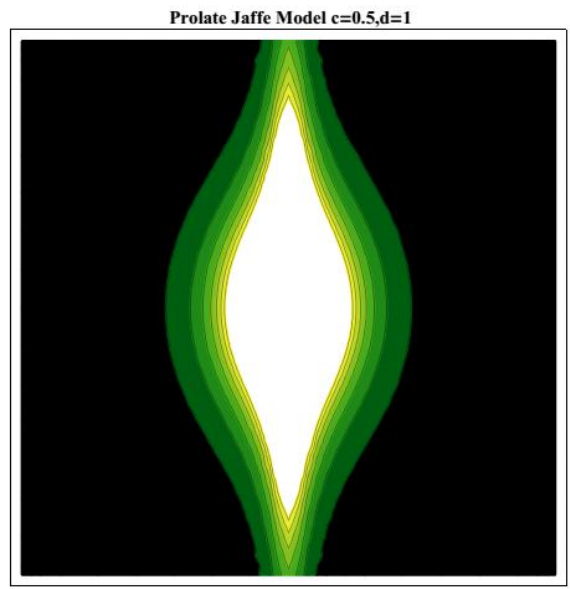

**Figure 9**: The density contours corresponding to Prolate Jaffe model in terms of different constants when the scale parameter is equal to 1.

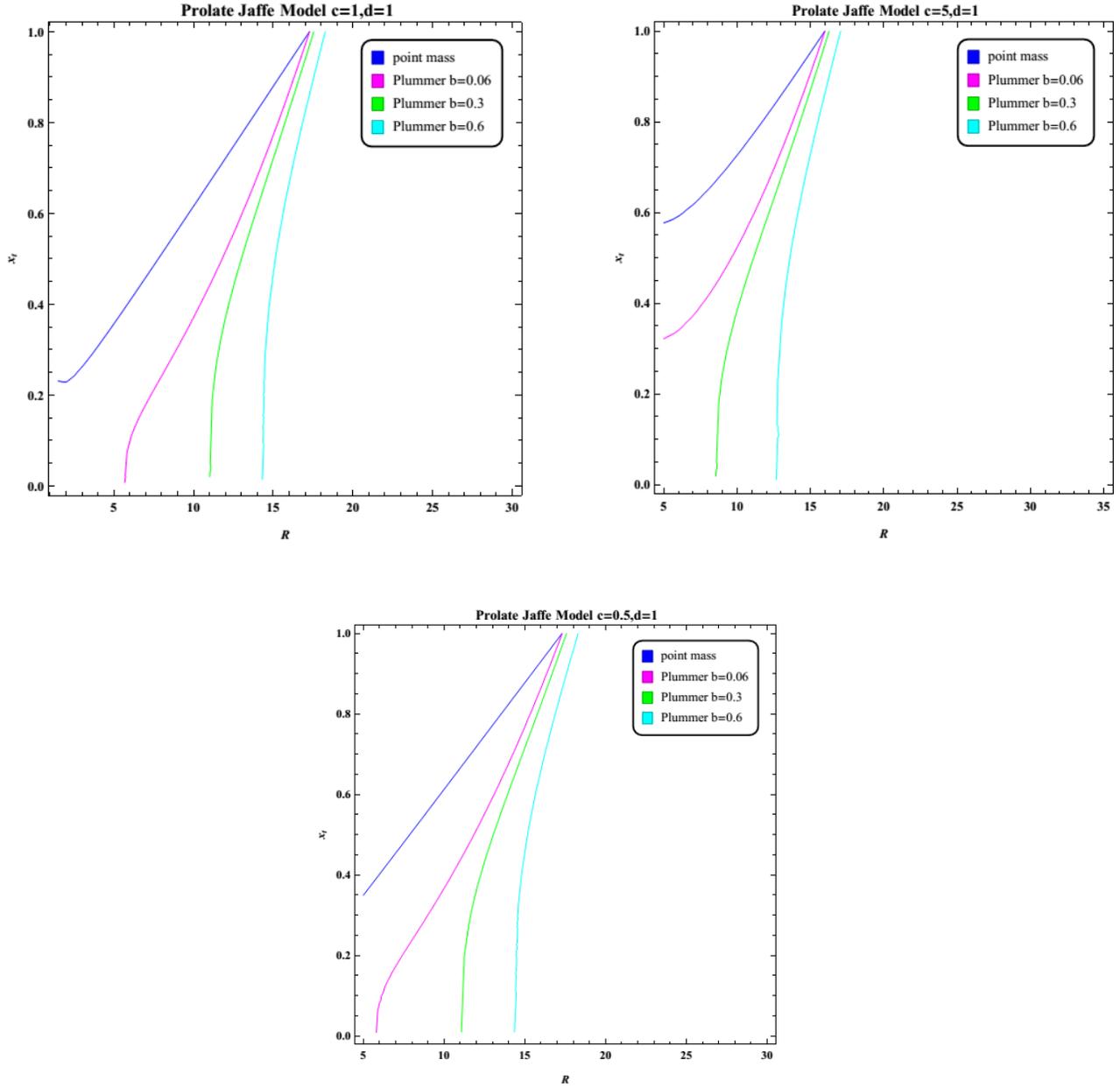

**Figure 10**: These curves show variation of tidal radius relative to the two systems distance. The galaxy model is based on the distributions shown in figure 9; the stellar system density is as the point mass and the Plummer models for scale lengths 0.06, 0.3, and 0.6.

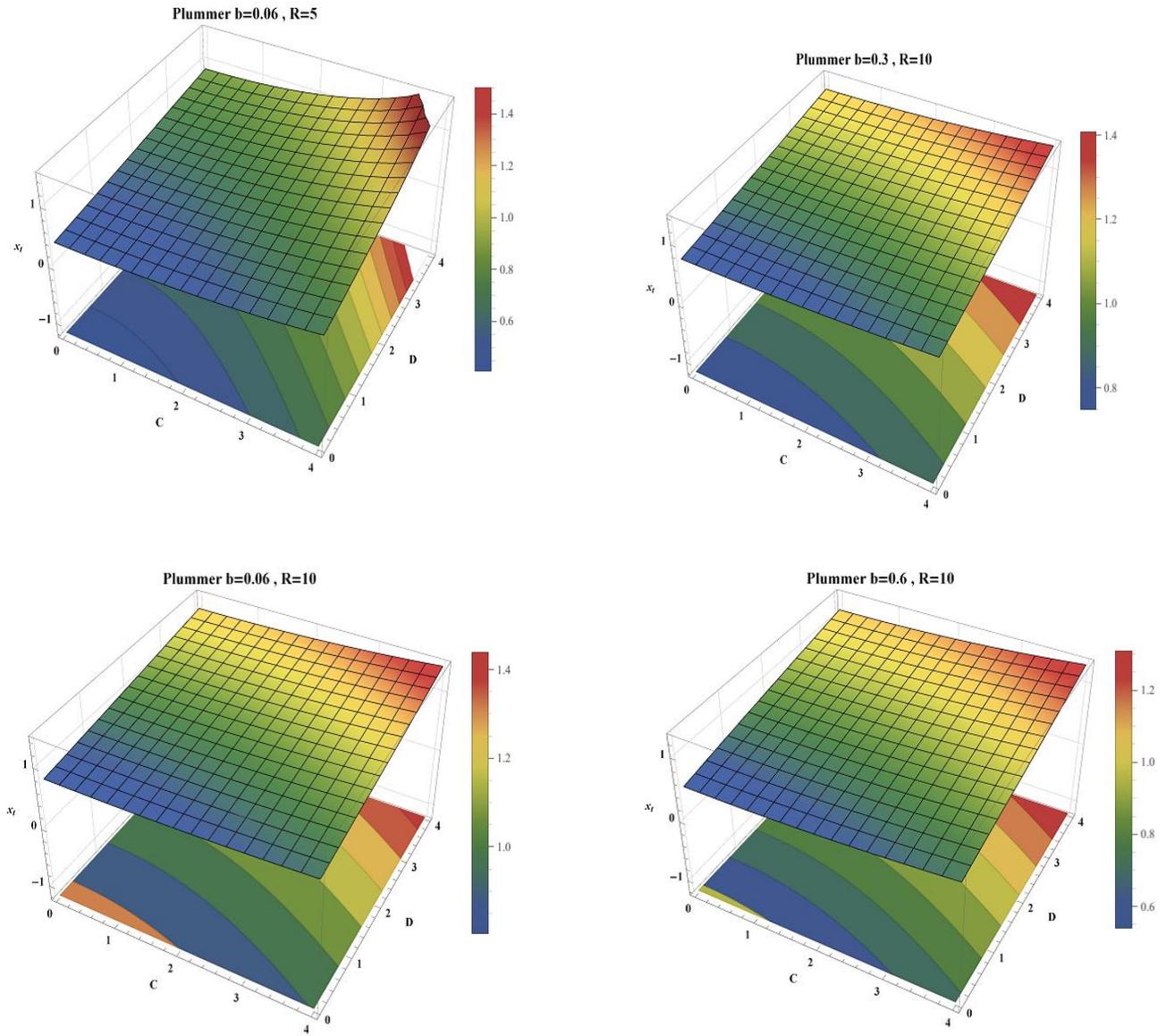

**Figure 11**: These curves show variation of tidal radius in terms of the two constants *C* and *D* (equation 25) for dimensionless distances 10 and 5. The galaxy model is a Prolate model and the stellar system density model is the Plummer one for scale lengths 0.06, 0.3, and 0.6.

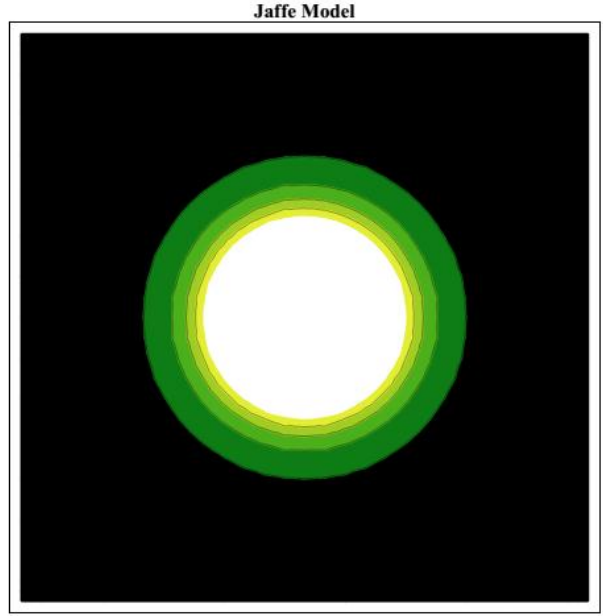

**Figure 12**: The density contours corresponding to the Jaffe model in terms of different constants when the scale parameter is equal to 1.

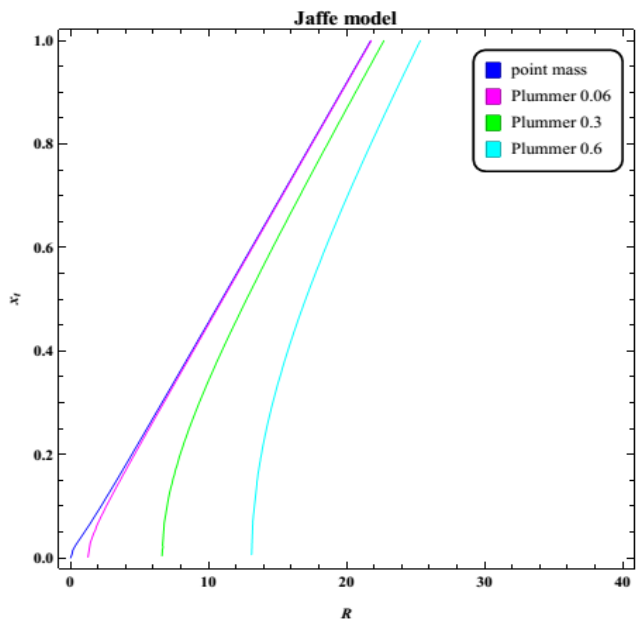

**Figure 13**: These curves show variation of tidal radius relative to the two systems distance. The galaxy model is based on the distributions shown in figure 12; the stellar system density is as the point mass and the Plummer models for scale lengths 0.06, 0.3, and 0.6.

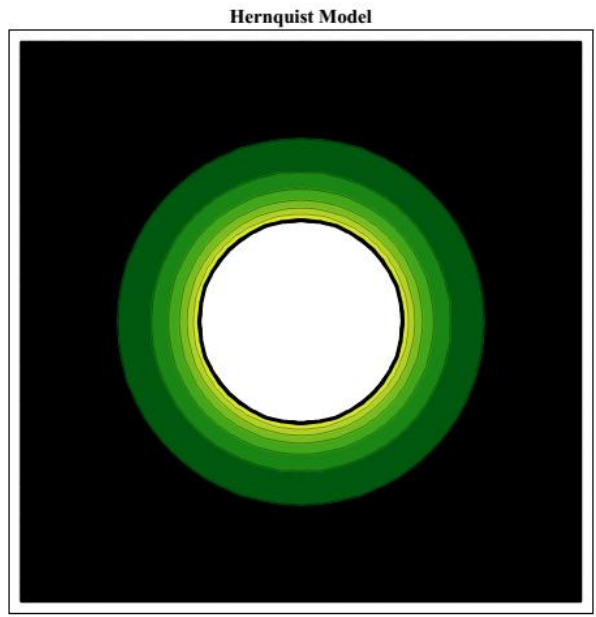

**Figure 14**: The density contours corresponding to the Hernquist model in terms of different constants when the scale parameter is equal to 1.

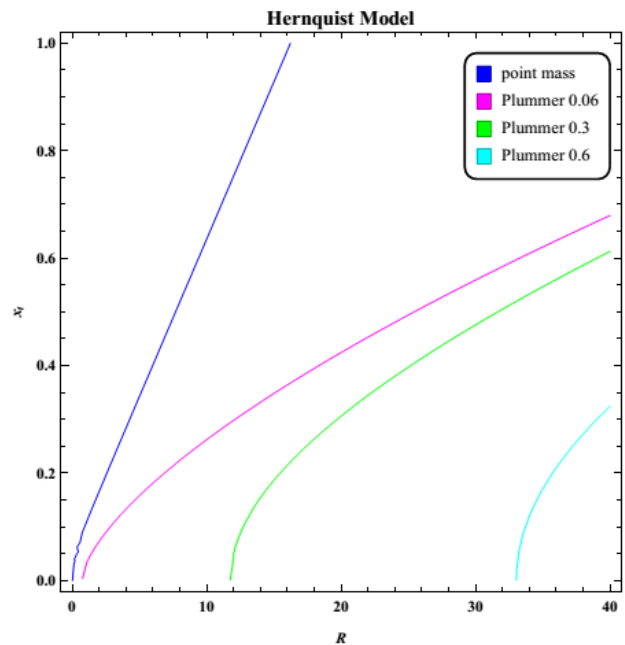

**Figure 15**: These curves show variation of tidal radius relative to the two systems distance. The galaxy model is based on the distributions shown in figure 14; the stellar system density is as the point mass and the Plummer models for scale lengths 0.06, 0.3, and 0.6.